\documentclass[prb,onecolumn,showpacs,preprintnumbers,amsmath,amssymb]{revtex4}

\usepackage{graphicx}
\usepackage{dcolumn}
\usepackage{bm}

\begin{document}

\newcommand{\up}{\uparrow}
\newcommand{\down}{\downarrow}
\newcommand{\phdagger}{\phantom{\dagger}}          
\newcommand{\aop}[1]{a^{\phdagger}_{#1}}           
\newcommand{\adop}[1]{a^{\dagger}_{#1}}            
\newcommand{\bop}[2]{b^{\phdagger}_{#1}(#2)}           
\newcommand{\bdop}[2]{b^{\dagger}_{#1}(#2)}            
\newcommand{\cop}[1]{c^{\phdagger}_{#1}}           
\newcommand{\cdop}[1]{c^{\dagger}_{#1}} 
\newcommand{\dop}[2]{d^{\phdagger}_{#1}(#2)}           
\newcommand{\ddop}[2]{d^{\dagger}_{#1}(#2)}            
\newcommand{\sS}{{\mathcal{S}}}
\newcommand{\sH}{{\mathcal{H}}}
\newcommand{\mgt}{>>}
\newcommand{\mlt}{<<}
\newcommand{\half}{\frac{1}{2}}
\newcommand{\rp}{r^{\prime}}
\def\C60{C$_{60}$}
\def\dbar{\bar{\Delta}}
\def\dbarbar{\bar{\bar{\Delta}}}
\def\spabcd{ { p_1 p_2 p_3 p_4 } }
\def\spcdef{ { p_3 p_4 p_5 p_6 } }
\def\spabef{ { p_1 p_2 p_5 p_6 } }
\def\ssabef{ { s_1 s_2, s_5, s_6}}
\def\sscdef{ { s_3 s_4 ,  s_5 s_6 } }
\def\ssabcd{ { s_1 s_2, s_3 s_4 } }
\def\sO{{\mathcal{O}}}
\def\cd{c^{\dagger}}
\newcommand{\bra}[1]{| #1 \rangle}


\title{Degenerate 3-band Hubbard model with anti-Hund's rule interactions; A model for A$_x$C$_{60}$.}
\author{Mats Granath}
\email{mgranath@fy.chalmers.se}
\affiliation{%
NORDITA\\
Blegdamsvej 17\\
DK-2100 Copenhagen \\
Denmark}
\author{Stellan \"Ostlund}%
\email{ostlund@fy.chalmers.se}
\affiliation{%
Chalmers Technical University\\
Gothenburg 41296 \\
Sweden
}


\date{\today}

\begin{abstract}

We consider the orbitally degenerate 3-band Hubbard model with on-site interactions which favor low spin 
and low orbital angular momentum using standard second order perturbation theory in the large Hubbard-U limit.
At even integer filling this model is a Mott insulator with a non-degenerate ground state that allows
for a simple description of particle-hole excitations as well as gapped spin and orbital modes. We find that the Mott
gap is generally indirect and that the 
single particle spectrum at low doping reappears close to even filling but rescaled by a factor $2/3$ or $1/3$. 
The model captures the basic phenomenology of the Mott insulating and metallic fullerides 
A$_x$C$_{60}$. This includes the existence of 
a smaller spin gap and larger charge gap at even integer filling, the fact that odd 
integer stoichiometries are generally metallic while even are insulating, as well as the rapid suppression of the 
density of states and superconducting transition temperatures with doping away from $x=3$.   

\end{abstract}

\maketitle
\section{Introduction}

The physics of Mott insulators has emerged as a key ingredient in the 
study of strongly correlated systems\cite{Sachdev}, largely motivated by the understanding that much of
the exotic physics of the underdoped cuprate superconductors has connections 
to the undoped Mott insulating state. The basic model for a Mott insulator 
discussed in connection with the cuprates is the large-U half-filled Hubbard model.
For this model the low 
energy spectrum at half-filling is reasonably well understood, but the higher energy spectrum as well as the physics 
away from half-filling
is still very much an open problem. Much of the difficulty seems to trace back to the
fact that the Mott insulating ground state for this system is not known. 

Here we will study a model of a Mott insulator which has a simple non-degenerate ground state and where we can 
describe the spectrum of excited states and explore the physics at and near the Mott transition in a well 
controlled manner. The model is a degenerate three channel Hubbard model which has multiplet splitting  
on-site interactions that favor low spin and low orbital angular momentum. 
This model is the simplest model with spin and orbital symmetry which at even integer filling
allows for a Mott insulating non-degenerate ground state.\cite{Mott_Note}
Since the ground state is non-magnetic, 
the spin physics which complicates the single band Hubbard model is absent at   
low energies.
Although interesting in itself as a natural and simplifying extension of the Hubbard model, our 
main motivation for studying this model 
is that it very naturally captures some of the most distinctive
phenomenology of the alkali doped \C60 compounds, the fullerides.   

Crystalline \C60 is a band insulator with 
a completely filled molecular orbital.\cite{Gunnarsson_review} Doping with alkali atoms, forming A$_n$\C60, 
transfers electrons into the 
lowest unoccupied molecular orbital (LUMO) which according to elementary molecular theory is 
threefold degenerate. These form three spin degenerate bands according to the band theory of the 
crystal. It is believed that the charge transfer is complete and that the influence
of the alkali ions is in general negligible apart from changing the crystal structure and the 
corresponding band theory. 

If these were simple metals, band theory would would predict  
a metallic state for any doping $0<n<6$ and possibly 
superconductivity with transition temperatures which would follow roughly the density of 
states at the Fermi energy as the doping is varied.  
Experimentally, however, it turns out that the compounds with even integer doping ($n=2,4$) are 
non-magnetic insulators\cite{opt,triplet_evidence,Kerkoud} 
and superconductivity is only seen in a narrow range around half-filling 
($n=3$) and with transition temperatures that are sharply maximized here\cite{Yildrim}. Band theory 
augmented by BCS theory of superconductivity fails to reproduce this behavior. 

It is well known that correlation effects are important as these are weakly bound molecular solids 
with a narrow band width of approximately $.5eV$ and it is believed that the on-site Coulomb repulsion
is approximately $1eV$. One might therefore expect that the system would be a 
Mott insulator at any integer filling. This, however, does not explain why  
systems with even integer filling are generally insulating while systems with odd integer filling are generally 
metallic. A second related question is 
why the insulating materials at even integer filling are non-magnetic. From Hund's rule, 
we would expect the highest spin 
configuration to be the molecular ground state, S=1 for n=2 and n=4, 
and consequently some sort of magnetic ground state is to be expected for the solid. 

One explanation for the violation of Hund's rule in the insulating materials is 
that the Jahn-Teller (JT) effect counteracts Hund's rule, 
with the lowest energy JT distorted state of the C$_{60}$ 
molecule at even integer occupation being a spin singlet.\cite{Auerbach} 
However, a problem with the naive Jahn-Teller scenario is that static distortions of the \C60 molecules 
have not been detected in solid \C60\cite{Kuntscher}.
Indirect evidence has come from the existence of two energy gap scales in the insulating systems,
a smaller spin gap of around $50meV$ and a larger charge gap of the around $500meV$. The large charge gap is 
quite clearly a Mott gap related to the intramolecular electron-electron repulsion, whereas the smaller spin gap 
has been linked to a singlet to triplet gap of JT distorted molecules.\cite{triplet_evidence}

A different scenario for the violation of Hund's rule suggested by 
Chakravarty et al.\cite{Chakravarty} and Baskaran and Tosatti\cite{Baskaran} claims that  
electronic correlations on a \C60 molecule can break the 
degeneracy of any partially filled molecular orbital in such a way as to minimize the spin and 
orbital angular momentum. Chakravarty and coworkers looked at a Hubbard model on single \C60 molecule, with 
strong electron-electron repulsion (U) on each carbon atom, using second order 
perturbation theory in U. The linear in U terms give Hund's rule with a preferred highest spin 
configuration which minimizes the overlap between electrons. The second order term on the other hand prefers 
low spin configurations with large overlaps which can take better advantage of virtual excitations of the 
core electrons.\cite{antiHund}      
The validity of the qualitative features of the perturbation theory is supported by exact diagonalization
of smaller Hubbard clusters.\cite{White}

One important feature of the scenario based on intramolecular electronic correlations is that in contrast to 
the static Jahn-Teller distortions the orbital symmetry of the molecule is preserved.  
More recently, it has also been emphasized in a series of papers by Tosatti and coworkers\cite{Fabrizio,CaponeI,CaponeII,CaponeIII} 
that Jahn-Teller phonons treated in the anti-adiabatic limit will also give rise to an effective electronic 
Hamiltonian with inverse Hund's rule which preserves the orbital symmetry of the molecule. It may thus be 
difficult to distinguish the latter scenario from the one based on electronic correlations and a full treatment
of both the electron-electron and the electron-lattice interactions to decide which energy scales dominate
is asked for. Along these lines a recent density-functional calculation 
does find that Hund's rule is valid for an isolated \C60 and only counteracted by the JT effect\cite{Luders}, although issues such as electronic screening 
by the surrounding molecules may also be important.\cite{Lemmert} 

Regardless of the mechanism behind the inversion of Hund's rule on the \C60 molecule it is an interesting problem to study
the effective model of solid \C60 which this naturally gives rise to, namely the three channel Hubbard model with ``anti-Hund's rule'' 
interactions. Here we study this model in the limit of strong interactions where we can do standard second order perturbation 
theory in the intermolecular hopping.
We find that the even/odd integer doping effect, the existence of two energy gaps in the insulating systems, 
as well as the rapid variation of T$_c$ with doping away from $n=3$ are all natural consequences of this model. 

We will focus primarily on even integer filling (n=2 or 4) where the problem simplifies significantly because of the resultant 
non-degenerate strong coupling ground state.
Based on this we can describe the spectrum consisting in general of an indirect Mott 
gap to particle-hole excitations and chargeless spin and orbital modes with a distinct energy gaps.
We also derive the spectrum of a single particle or hole doped into the non-magnetic Mott insulator. 
The particle and hole spectra turns out to be exactly the
same as the non-interacting band structure up to an overall rescaling of the hopping by a factor $1/3$ or
$2/3$. These characteristic properties of the non-magnetic Mott insulator are in agreement with previous results on the same model.\cite{Fabrizio,CaponeI}

Apart from the correspondence with the physics of the Mott insulating A$_2$\C60 and A$_4$\C60 
our main conclusion is that 
C$_{60}$ with a filling $n=2+x$ or $4+x$ of the LUMO band, with $|x|\mlt 1$, is best understood as a
{\em doped Mott insulator},
manifested by a charge carrier concentration given by $|x|$ 
instead of $n$. In addition the density of states is up to a rescaling factor the same as that of the 
non-interacting band structure at the band edges $n=|x|$ or $n=6-|x|$ for particle and hole doping respectively.
If we heuristically extend these 
results to larger $x\approx 1$ it follows 
naturally the observed {\em variation} of T$_c$ with doping for $2<n<4$ within any weak coupling BCS like 
scenario as a consequence of the rapid drop in the density of states close to non-interacting band edges.  

We will not, however, discuss the actual mechanism of superconductivity, i.e. the source of
pairing. In particular, we will not assume that the intramolecular singlet triplet gap is necessarily 
large enough to overcome the Coulomb repulsion and give rise to a bare attraction along the lines of the 
earlier work.\cite{Chakravarty,Baskaran}
In fact, a fit of the parameters of the model to experiment
suggests that this is not the case. Nevertheless, any other
mechanism, such as one based on attraction from electron-phonons interactions, certainly needs 
to include the strong electron correlations which are present as indicated most clearly by existence
of Mott insulating phases.

The paper is organized as follows: In Sec. \ref{sec:Model} we define the model and the strong coupling limit
we will primarily consider. Then we derive a perturbative effective Hamiltonian valid in the strong coupling limit.
In Sec. \ref{sec:Mott_insulator} we study the Mott insulator at even integer filling and describe the spectrum of
excited states. We also discuss the distinction between even and odd integer doping and why at odd integer doping
the system is more likely to be on the metallic side of the Mott transition. Then we
use the results derived for the model to get an estimate of the parameters by comparing to 
experiment on A$_2$\C60 and A$_4$\C60. In Sec. \ref{2+x} we study the doped Mott insulator at 
a filling close to even integer and speculate on the implications of these results to the whole doping range
$2<n<4$ as well as the properties at higher temperatures. Finally, in Sec. \ref{Summary} we conclude.

\section{\label{sec:Model} The Model}

We will be studying a three band Hubbard model. The electrons occupy three degenerate p-orbitals ($L=1$) on 
every site of the lattice and we define electron creation and destruction operators $\cdop{r,ls}$ and 
$\cop{r,ls}$ with site index, orbital quantum number and spin respectively. We will be working in the $L^z$ basis
where $l=-1,0,1$. 
 
The Hamiltonian reads
\begin{equation}
H =  h + H_I  \label{theHamiltonian}
\end{equation}
with a nearest neighbor hopping
\begin{equation}  
h = \sum_{\langle rr'\rangle}t^{rr'}_{ll'}\cdop{r,ls}\cop{r',l's}\,,
\end{equation}
and on-site interaction
\begin{equation}
H_I =\sum_{r} \half U n_{r}^2 +J_L\vec{L}_{r}^2+ J_S \vec{S}_{r}^2\,,\label{H_I}
\end{equation}
with $U,J_L,J_S>0$.
Here 
\begin{eqnarray}
n_{r} = &  \sum_{l,s} \cdop{r,ls}\cop{r,ls}\\
\vec{L}_{r} =&\sum_{ll's}\cdop{r,ls} \vec{L}_{l,l'}\cop{r,l's}\\
\vec{S}_{r} =   & \sum_{lss'}\cdop{r,ls}\vec{S}_{s,s'}\cop{r,ls'}\,.
\end{eqnarray}
are the number, orbital angular momentum and spin operators respectively.

The lattice is three dimensional and we assume a point group symmetry which is such that the hopping preserves the
threefold orbital degeneracy, although this may be relaxed as long as the symmetry breaking crystal field is small
compared to the interactions.  
We are going to be studying this in the strong coupling limit defined as 
\begin{equation}
U \mgt J_S \mgt J_L\mgt t\,,\label{strong_coupling} 
\end{equation}
where $t=\text{max}(t^{rr'}_{ll'})$. For \C60 band structure calculations\cite{Laouini} give $t\sim 100meV$ and 
values of $U$ of the order $1eV$ have been suggested\cite{Gunnarsson_review}. 
A subsequent fit of the parameters of the model to experiment on A$_n$\C60 suggests that the large U limit is 
quite well satisfied with $U/t\approx 5$, while $J_S\sim J_L\sim t$. The limit $U\mgt t$ is essential for our 
treatment while relaxing the other limits probably will not change the qualitative features of our results. 

At $t=0$ we find that the Hamiltonian is simply diagonalized in terms of the states $\bra{n,L,S,L^z,S^z}_r$ with $n$
the number of electrons on a site and $(L,S)$ the total angular momentum and 
spin and the energy $E(n,L,S)=\half Un^2+J_LL(L+1)+J_SS(S+1)$. In the strong coupling limit the effect of the 
hopping $t$ will be to reduce the translational degeneracy.
It is the topic of the next section to derive an effective Hamiltonian which describes these low energy degrees of
freedom.

\begin{table}
\begin{center}
\begin{tabular}{|c|c|c|} \hline
{\bf n} & {\bf (L,S)} & E\\ \hline\hline
0& $(0,0)$ & 0 \\ \hline
1& $(1,\half)$ & $\half U+2J_L+\frac{3}{4}J_S$\\ \hline
2& $(0,0)$,\,$(2,0)$,\,$(1,1)$ & $2U+\{0,\,6J_L,\,2J_L+2J_S\}$\\ \hline
3& $(1,\half)$,\,$(2,\half)$,\,$(0,\frac{3}{2})$ & 
$\frac{9}{2}U+\{2J_L+\frac{3}{4}J_S,\,6J_L+\frac{3}{4}J_S,\,\frac{15}{4}J_S\}$\\ \hline
4& $(0,0)$,\,$(2,0)$,\,$(1,1)$ & $8U+\{0,\,6J_L,\,2J_L+2J_S\}$\\ \hline
5& $(1,\half)$ & $\frac{25}{2}U+\frac{3}{4}J_S$\\ \hline
6& $(0,0)$ & 18U\\ \hline
\end{tabular}
\caption{Spectrum of the interaction, Eq. \ref{H_I}, at a single site with states specified by 
occupation $n$, total orbital angular momentum $L$ and total spin $S$.\label{t0spectrum}}
\end{center}
\end{table}

\subsection{\label{sec:level2} Effective strong coupling Hamiltonian.}

Let us define the short-hand notation $p=(n,L,S)$ and $\alpha=(L^z,S^z)$. As shown in 
Table \ref{t0spectrum} any single site states 
$|p,\alpha\rangle_r$ and $|p',\alpha'\rangle_r$ have the same energy, barring accidental degeneracies, 
if and only if $p=p'$.
This implies that any eigenstate of $H_I$ will not mix different representations at a site
and can be written as $\prod_r\psi_{p,\alpha}^r|p,\alpha\rangle_r$, 
where $\psi_{p,\alpha}^r\neq 0$ only for one particular $p(r)$.
Consider now the action of the hopping term $h$ on an eigenstate. This will affect two arbitrary 
nearest neighbor sites $r$ and $r'$, giving  
$|p,\alpha\rangle_r|p',\alpha'\rangle_{r'}\overset{t}{\rightarrow}
\sum_{q\beta q'\beta'}M|q,\beta\rangle_r|q',\beta'\rangle_{r'}$, where M are the matrix 
elements of h between the in and out states.
Unless the initial and final nearest 
neighbor states have only exchanged particle number and representation between the sites, i.e. 
$q=p'$ and $q'=p$, there 
will be a non-zero energy difference $\Delta E=E(q)+E(q')-E(p)-E(p')$ 
with respect to $H_I$ which is some linear 
combination of $U,J_L,J_S$. In the strong coupling limit Eq. \ref{strong_coupling} there is no
accidental degeneracy and the 
energy difference satisfies $\Delta E\mgt t$, which will allow us to do a perturbation theory in 
$t/\Delta E$. Henceforth we will use the $\Delta E$ to define the absolute value of the minimum 
non-zero energy difference between
two different nearest neighbor state configurations. From table \ref{t0spectrum} we can check that 
in the strong coupling limit this corresponds to $\Delta E=4J_L$, given for instance by 
$E(3,2,\half)+E(2,0,0)-E(2,0,0)-E(3,1,\half)$.

We will be using the Effective Hamiltonian method described as applied to the one band Hubbard model in, for 
instance, Fazekas\cite{Fazekas}. 
We introduce a canonical transformation 
\begin{equation}
\sH_{eff}  = e^{i \sS } \sH e^{-i \sS }  = \sH + i [\sS,\sH] + \frac{i^2}{2} 
[\sS,[\sS,H]] + ...\,,\label{canon}
\end{equation}
where the goal is to find $\sH_{eff}$ such that it does not couple the different energy sectors of $H_I$.

Let us consider $h$ as a matrix which acts on nearest neighbor states 
$|p,\alpha\rangle_r|p',\alpha'\rangle_{r'}=|a\rangle_{rr'}$ which we in a short-hand notation will 
denote by a single index $a,b,c,...$. We write 
\begin{equation}
h=\sum_{\langle rr'\rangle}h^{rr'}=\sum_{\langle rr'\rangle}h^{rr'}_{ab}|a\rangle_{rr'}\langle b|_{rr'}\,,
\end{equation}
where 
\begin{equation}
h^{rr'}_{ab}=\langle a|_{rr'}\sum_{ll'}t^{rr'}_{ll'}(\cdop{r,ls}\cop{r',l's}+\text{h.c.})
|b\rangle_{rr'}\,.\label{hrr}
\end{equation}
Next we split $h^{rr'}_{ab}$ into $h^{rr'}_{ab}=h^{0,rr'}_{ab}+h^{1,rr'}_{ab}$ where
$h^{0,rr'}_{ab}=h^{rr'}_{ab}
\delta_{E(a),E(b)}$, 
only connects states with the same energy and 
$h^{1,rr'}_{ab}=h^{rr'}_{ab}-h^{0,rr'}_{ab}$ which
connects states with different energy. It follows that we can also write $h=h^0+h^1$.

We introduce $ \sS=\sS_1+\sS_2$ where $\sS_1\sim\sO(t/\Delta E)$ and 
$\sS_2\sim\sO(t^2/\Delta E^2 )$ and expand Eq. \ref{canon} to 
$\sO ( t^2 /\Delta E  )$,
\begin{equation}
\sH_{eff} = H_I+ h^0 + h^1 + i [ \sS_1, H_I ] + i [ \sS_1, h^0 + h^1 ]+
\frac{i^2}{2}  [ \sS_1, [\sS_1, H_I ]] + i [ \sS_2, H_I ]\,. \label{seff}
\end{equation} 
To first order in $t$ we want to cancel $h^1$ by taking  
$\sS_1$ to solve  
\begin{equation}  
i [ \sS_1, H_I ] =-h^1 \label{eq:sol}
\end{equation}

Clearly $\sS_1$ needs to act only on nearest neighbor sites. Defining 
$\sS_1=\sum_{\langle rr'\rangle}\sS_1^{rr'}=\sum_{\langle rr'\rangle}\sS_{1,ab}^{rr'}|a\rangle_{rr'}\langle b|_{rr'}$ 
gives from Equation (\ref{eq:sol}) 
\begin{eqnarray}
i\sum_{\langle rr'\rangle}h^{1,rr'}&=&\sum_{\langle rr'\rangle r''}[\sS_1^{rr'},H_I^{r''}]
\nonumber\\
&=&\sum_{\langle rr'\rangle}\sS_{1,ab}^{rr'}[|a\rangle_{rr'}\langle b|_{rr'},H_I^r+H_I^{r'}]\nonumber\\
&=&\sum_{\langle rr'\rangle}\sS_{1,ab}^{rr'}(E(b)-E(a))|a\rangle_{rr'}\langle b|_{rr'}\,,\label{eq2forS}
\end{eqnarray}
where we have used the fact that we work in the eigenbasis of $H_I$ such that 
$(H_I^r+H_I^{r'})|a\rangle_{rr'}=E(a)|a\rangle_{rr'}$. We thus arrive at the solution 
\begin{equation}
S^{rr'}_{1,ab}=\frac{ih^{1,rr'}_{ab}}{E(b)-E(a)}\,.
\end{equation} 
Recall that $h^{1,rr'}_{ab}$ is defined as to only have non-zero entries when 
$|E(b)-E(a)|\geq\Delta E$ so that $\sS_1\sim t/\Delta E$.

Given the solution for $\sS_1$ obeying Eq. \ref{eq:sol} we can rewrite Eq. \ref{seff} as
\begin{eqnarray}
\sH_{eff}  & =  H_I+h^0+i[\sS_1,h^0+h^1]-\frac{i}{2}[\sS_1,h^1]+i[\sS_2,H_I]\nonumber\\
	   & =  H_I+h^0+i[\sS_1,h^0]+\frac{i}{2}[\sS_1,h^1]+i[\sS_2,H_I]\label{Heff1}
\end{eqnarray}

In analogy to using $\sS_1$ to cancel terms linear in $t$ which connect different energy subsectors
we can now use $\sS_2$ to cancel similar terms to order $t^2$ which arise from the commutators of 
$\sS_1$ and $h$ in Eq. \ref{Heff1}. Technically this is slightly more involved because 
there is also three site next nearest neighbor interactions generated to second order in 
$t$ and we refer the reader to the 
Appendix for the details. 

The final expression for the effective strong coupling Hamiltonian reads
\begin{equation}
\sH_{eff}=H_I+\sum_{\langle r,r'\rangle,ab}(\sH_{eff})^{rr'}_{ab}|a\rangle_{rr'}\langle b|_{rr'}
+\sum_{\langle r,r'r''\rangle,ab}(\sH_{eff})^{rr'r''}_{ab}|a\rangle_{rr'r''}\langle b|_{rr'r''}
+\sO(t^3/\Delta E^2)\,,
\end{equation}
with
\begin{eqnarray}
\label{Hefffinal}
(\sH_{eff})^{rr'}_{ab}&=&\left(
h^{rr'}_{ab}-\sum_{c,E(c)\neq E(a)}\frac{h^{rr'}_{ac}
h^{rr'}_{cb}}{E(c)-E(a)}\right)\delta_{E(a),E(b)}\\
(\sH_{eff})^{rr'r''}_{ab}&=&
-\half\sum_{c,E(c)\neq E(a)}\frac{h^{rr'}_{ac}
h^{r'r''}_{cb}+h^{r'r''}_{ac}h^{rr'}_{cb}}{E(c)-E(a)}\delta_{E(a),E(b)}\,.
\end{eqnarray}
Here $|a\rangle_{rr'}=|n,L,S,L^z,S^z\rangle_r|n',L',S',{L'}^z,{S'}^z\rangle_{r'}$ and $|a\rangle_{rr'r''}$ 
are arbitrary nearest neighbor and next nearest neighbor eigenstates of $H_I$ with energy $E(a)$ and 
$h^{rr'}_{ab}$ is defined according to Eq. \ref{hrr} with a straightforward extension for three site states.

\section{The Mott insulator\label{sec:Mott_insulator}}

In the strong coupling limit, Eq. \ref{strong_coupling}, of this model the system is an insulator at any integer 
filling with a gap to charge carrying excitations of order $U$. At odd integer filling the ground state 
is the highly degenerate $|GS,1/3/5\rangle=\prod_{r}|n=1/3/5,L=1,S=\half,L^z,S^z\rangle_r$ and the hopping $t$
will introduce strong correlations in analogy with the half-filled single band Hubbard model. 
We will return to this problem briefly in discussing the fact that A$_3$\C60 is generally on the metallic side 
of the Mott transition.

However, at even integer filling, $n=2$ or 4, the ground state is non-degenerate, 
$\bra{GS,2/4}=\prod_r\bra{n=2/4,L=0,S=0}_r$. Here, we will find that much of the physics can be understood in terms
of a simple non-interacting single particle picture. This will be the focus of the subsequent discussion and the 
main issue of the paper. 
We will discuss $n=4$ to compare with experiments on K$_4$\C60 and 
Rb$_4$\C60 but the treatment of $n=2$ is completely analogous. 

\begin{figure}[h]
\includegraphics{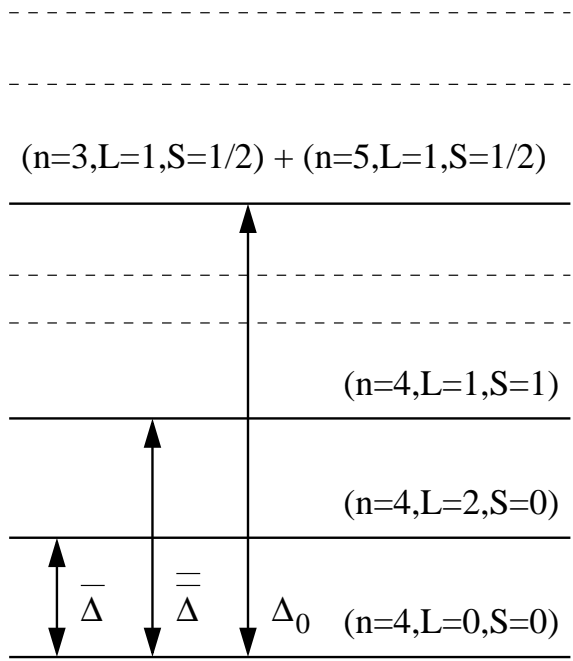}
\caption{\label{fig:mottfig1} Spectrum at $t=0$ and doping $n=4$ showing the lowest energy excitations 
of the three distinct kinds discussed in the text.}
\end{figure}

At filling $n=4$ there are three types of excitations of the ground state that can be readily identified and 
which are indicated in 
figure \ref{fig:mottfig1}. We can excite a single site into a higher energy multiplet, creating states
$|4,2,0,L^z\rangle_r\prod_{r'\neq r}\bra{4,0,0}_{r'}$ or 
$|4,1,1,L^z,S^z\rangle_r\prod_{r'\neq r}\bra{4,0,0}_{r'}$ with energy 
$\dbar\equiv E(4,2,0)-E(4,0,0)=6J_L$ and $\dbarbar\equiv E(4,1,1)-E(4,0,0)=2J_L+2J_S$ respectively.
There is also ``particle-hole'' excitations in different multiplets, the one with lowest energy being
$|3,1,\half,L^z,S^z\rangle_r|5,1,\half,{L'}^z,{S'}^z\rangle_{r'}
\prod_{r''\neq r,r'}\bra{4,0,0}_{r''}$ with energy $\Delta_0=U+4J_L+\frac{3}{2}J_S$. The energy $\Delta_0$
is the Mott gap to 0'th order in $t$.

The degeneracy of the excited states will be lifted by the hopping $h$ and using the effective 
Hamiltonian Eq. \ref{Hefffinal} we can study the spectrum perturbatively in $t$.

\subsection{Spin gap, spin and orbital modes\label{sec:spin_modes}}

The spin and orbital excitations $|4,1,1,L^z,S^z\rangle_r\prod_{r'\neq r}\bra{4,0,0}_{r'}$
and $|4,2,0,L^z\rangle_r\prod_{r'\neq r}\bra{4,0,0}_{r'}$ are degenerate in $L^z$ and $S^z$ as well as position $r$.
Acting with the effective Hamiltonian Eq. \ref{Hefffinal} will split the degeneracy in space (as well as the orbital 
degenercy) leading to a band 
description of these states. Clearly to first order in $h$ ($t$), the effective Hamiltonian will not affect
these states as it necessarily creates a high energy particle-hole state. For the same reason the second order
three site interaction will not contribute. 
However, the second order nearest neighbor term can hop the excited state at $r$ to a nearest neighbor 
$r'$ or to the same site $r$ through an intermediate particle-hole state. 

Focusing on the spinless excitations we define the ``bosonic'' operator 
\begin{equation}
b^{\dagger}_{r,m}=|4,2,0,m\rangle_r\langle 4,0,0|_r
\end{equation}
in terms of which we can write an excited orbital state as $b^{\dagger}_{r,m}|GS,4\rangle$. These operators do not
obey proper commutation relations but if we consider only single particle 
physics this is irrelevant. With these operators we can write the following single particle
Hamiltonian describing the dynamics of such excitations
\begin{equation}
H=\sum_{\langle r,r'\rangle}\bar{t}^{rr'}_{mm'}b^{\dagger}_{r,m}b_{r',m'}
+\sum_{r}\bar{t}_{mm'}b^{\dagger}_{r,m}b_{r,m'}\,,\label{orbital_hopping}
\end{equation}   
where
\begin{eqnarray}
\bar{t}^{rr'}_{mm'}=-\frac{1}{U+3/2J_S-2J_L}\sum&&\langle 4,2,0,m|_r\langle 4,0,0|_{r'}t^{rr'}_{ii'}
(\cdop{r,is}\cop{r',i's}+\text{h.c.}) 
|\text{ph}\rangle\times\nonumber\\
&&\langle\text{ph}| t^{rr'}_{jj'}(\cdop{r,js}\cop{r',j's}+\text{h.c.})|4,0,0\rangle_r|4,2,0,m'\rangle_{r'}\,,
\label{t_boson}
\end{eqnarray}
and similarly for $\bar{t}_{mm'}$. Here $|\text{ph}\rangle$ are particle-hole states 
$|5,1,\half,l,s\rangle_r|3,1,\half,l',s\rangle_{r'}$ (or with reversed positions) and the sum is over all indices except
$m,m'$ and $r,r'$. Similar 
contributions but with larger denominators from intermediate particle hole excitations in higher energy multiplets 
have been ignored in Eq. \ref{t_boson}. 

Deriving the Hamiltonian Eq. \ref{orbital_hopping} is thus a straightforward problem.
The Hamiltonian is non-interacting and can in principle be diagonalized in momentum space. 
In general the five orbital
components will be mixed with a complicated band structure. The same calculation can be carried out for the 
spinful $(n=4,L=1,S=1)$ states, where there are nine coupled states per site. 
Although the band width of these spin and orbital modes are naively order $t^2/U$ the high degeneracy of 
these states is expected to give a significant broadening of band width, perhaps by a factor of the spin and orbital
degeneracy of these states. The spinful excitations should be readily detectable experimentally as a band of 
magnons with some gap $\Delta_s$, a spin gap. We will return to the issue of the spin modes in discussing NMR on 
A$_2$\C60 and A$_4$\C60. 
  
\subsection{Charge Excitations}

At filling $n=4$ the lowest energy charge carrying excitations are the ``particle-hole'' states 
$|3,1,\half,l,s\rangle_r|5,1,\half,l',s'\rangle_{r'}
\prod_{r''\neq r,r'}\bra{4,0,0}_{r''}$ with an energy $\Delta_0=U+4J_L+\frac{3}{2}J_S$. 
Acting with the 
effective Hamiltonian Eq. \ref{Hefffinal} on such a state will translate either the particle ($n=5$) or the
hole ($n=3$) to a nearest neighbor site through the linear term in $h$ ($t$). To second order, there is next nearest
neighbor hopping through an intermediate state in a higher energy multiplet. Since it is higher order, 
$\sO(t^2/\Delta E)$, 
we will ignore this contribution. However, it should be noted that the denominator $\Delta E$ for this process is 
proportional to $J_S$ and/or $J_L$ not $U$, implying of course that we should consider the higher energy particle 
and hole multiplets if the strong coupling limit Eq. \ref{strong_coupling} is not strictly valid.   
Note that due to the energy constraint in the effective Hamiltonian the particle and hole are not 
allowed to annihilate by 
hopping to the same site. We will ignore the constraint which in three dimensions is expected to give a vanishing 
contribution to the spectrum at low densities. 

\subsubsection{Charged single particle excitations\label{single_particle_section}}
In order to understand charge transport, we need to consider the single particle and single hole states 
$|5,l,s\rangle_r\prod_{r'}|4\rangle_{r'\neq r}$ and $|3,l,s\rangle_r\prod_{r'}|4\rangle_{r'\neq r}$. 
Here we have dropped the multiplet indices $(L,S)=(1,\half)$ or $(0,0)$.
When acted on by $\sH_{eff}$ these excitations are translated to a nearest neighbor through $h$
with a matrix element
\begin{eqnarray}
\langle 4|_r\langle 5,l',s'|_{r'}(\sum_{jj',\sigma}t^{rr'}_{j'j}c^{\dagger}_{r',j',\sigma}c_{r,j,\sigma})
|5,l,s\rangle_r|4\rangle_{r'}&=&
\frac{1}{3}\delta_{ss'}t^{rr'}_{l'l}\nonumber\\
\langle 4|_r\langle 3,l',s'|_{r'}(\sum_{jj',\sigma}t^{rr'}_{j'j}c^{\dagger}_{r,j,\sigma}c_{r',j',\sigma})
|3,l,s\rangle_r|4\rangle_{r'}&=&
-\frac{2}{3}\delta_{ss'}t^{rr'}_{l'l}\,,\label{hopping_integrals}
\end{eqnarray}
and similarly with a factor $-1/3$ for holes and $2/3$ for particles with respect to the $n=2$ ground state.
The matrix elements in Eq. \ref{hopping_integrals} follow from the explicit expressions for the states
\begin{eqnarray}
|5,1,\half,l,s\rangle &=&\sqrt{3}\cdop{ls}|4,0,0\rangle\nonumber\\
|3,1,\half,l,s\rangle &=&\sqrt{\frac{3}{2}}\cdop{ls}|2,0,0\rangle=
\sqrt{\frac{3}{2}}(2s)^{2|l|-1}\cop{-l-s}|4,0,0\rangle\nonumber\\
|1,1,\half,l,s\rangle &=&\cdop{ls}|0\rangle=\sqrt{3}(2s)^{2|l|-1}\cop{-l-s}|2,0,0\rangle\,,
\end{eqnarray}
which is a simple exercise in elementary quantum mechanics to derive.

We note the reduced magnitude of the matrix elements compared to the single particle particle hopping on empty sites
\begin{equation}
\langle 0|_r\langle 1,l',s|_{r'}(\sum_{jj',\sigma}t^{rr'}_{j'j}c^{\dagger}_{r',j',\sigma}c_{r,j,\sigma})
|1,l,s'\rangle_r|0\rangle_{r'}=\delta_{ss'}t^{rr'}_{l'l}\,.
\end{equation}
This is a general result of the reduced Hilbert space due to constraining the $n$ particle states to the 
lowest energy multiplet.

Let us define the particle and hole creation operators 
\begin{eqnarray}
c^{\dagger}_{5,rls}&=&|5,l,s\rangle_r\langle 4|_r\nonumber\\
c^{\dagger}_{3,rls}&=&|3,l,s\rangle_r\langle 4|_r\,,\label{ph_ops}
\end{eqnarray}
through which the particle and hole states can be written as $c^{\dagger}_{5,rls}|GS,4\rangle$ and 
$c^{\dagger}_{3,rls}|GS,4\rangle$ with $|GS,4\rangle=\prod_{r}|4,0,0>_r$. 
Note that these operators do not obey on-site anticommutation relations, we can only use them with 
confidence in describing non-interacting single-particle physics. 

In terms of these particle and hole operators we now straightforwardly arrive at the 
following single particle Hamiltonians for the particle and hole states 
\begin{eqnarray}
H_{5}=\sum_{\langle rr'\rangle,ll's}\frac{1}{3}t^{rr'}_{ll'}c^{\dagger}_{5,rls}c_{5,r'l's}+
\sum_{r,ls}(9/2U+3/4J_S+2J_L-\mu)c^{\dagger}_{5,rls}c_{5,rls}\label{5sp}\\
H_{3}=-\sum_{\langle rr'\rangle,ll's}\frac{2}{3}t^{rr'}_{ll'}c^{\dagger}_{3,rls}c_{3,r'l's}+
\sum_{r,ls}(-7/2U+3/4J_S+2J_L+\mu)c^{\dagger}_{3,rls}c_{3,rls}\,,\label{3sp}
\end{eqnarray}
where the on-site energy is defined with respect to the $E(4,0,0)=8U-4\mu$ and where we have introduced the 
chemical potential $\mu$.

These are just simple tight binding Hamiltonians which we can diagonalize in momentum space 
in terms of states
\begin{eqnarray}
\bra{5,k,ls}=\cdop{5,kls}\bra{GS,4}=\frac{1}{\sqrt{V}}\sum_r e^{i\vec{k}\cdot\vec{r}}\cdop{5,rls}\bra{GS,4}\nonumber\\
\bra{3,k,ls}=\cdop{3,kls}\bra{GS,4}=\frac{1}{\sqrt{V}}\sum_r e^{-i\vec{k}\cdot\vec{r}}\cdop{3,rls}\bra{GS}.
\label{k_space_ph} 
\end{eqnarray}
In general these are not eigenstates due to the fact that the hopping is not diagonal in $l$ 
and we would get three (or more if there are inequivalent sites) 
non-degenerate bands depending on the precise nature of the 
hopping integrals which depend on the crystal symmetry.
Quite intriguingly, as seen from Equations \ref{5sp} and \ref{3sp}, 
the band structure for these particle and hole excitations from the Mott insulating 
ground state is precisely the noninteracting band structure up to a rescaling factor. The reason for this 
remarkably simple behavior is that the ground state at even integer filling is in the trivial $(L=0,S=0)$ 
representations of spin and angular momentum and as such is basically equivalent to the zero-particle vacuum. 

\subsubsection{Mott gap}

The Mott gap is defined as the gap between the ground state and the lowest energy charge carrying 
excitation of the insulator. What is often measured however is the optical gap as defined by 
optical conductivity or reflectivity measurements and we will be interested in calculating this too.

The Kubo formula for the optical conductivity which is the short wavelength limit of the 
electrical conductivity is at zero temperature 
\begin{equation}
\sigma_{aa}(\omega,q=0,T=0)=\frac{2i\hbar^2e^2}{m^2V}\sum_m\frac{\omega}{\omega_m}
\frac{|\langle m|j^a(q=0)|GS\rangle|^2}{\omega(\omega+i\eta)-\omega_m^2}\,\label{optcond}
\end{equation}
where $|m\rangle$ are excited states with $\hbar\omega_m=E_m-E_{GS}$.\cite{Mahan}
The current operator can be derived from the continuity equation $i\hbar^{-1}[n(r),H]+\nabla\cdot\vec{j}(r)$.
For the Hamiltonian Eq. \ref{theHamiltonian} considered here only the tight binding part contributes and we get
\begin{equation}
\vec{j}(\vec{q})=\sum_{\vec{p},ll's}\left(\frac{\partial}{\hbar\partial\vec{p}}\sum_{\delta}
e^{-i\vec{p}\cdot\vec{\delta}}t^{0\,\delta}_{ll'}
\right)\,\cdop{p+q,ls}\cop{p,l's}\,,
\end{equation}
where $\vec{\delta}$ is the set of nearest neighbor lattice vectors.
For a cubic or orthorhombic lattice this simplifies to
\begin{equation}
\vec{j}(\vec{q})=-\frac{1}{\hbar}\sum_{\vec{p},ll's,\delta}
\delta^a\sin(\vec{p}\cdot\vec{\delta})\,t^{0\,\delta}_{ll'}\,\cdop{p+q,ls}\cop{p,l's}\,.
\end{equation}

The current operator creates particle-hole pairs when acting on the ground state $|GS,4\rangle$ 
and we expect to get non-zero
matrix elements with states 
\begin{equation}
\bra{kls,k'l's'}_{\text{p-h}}=\cdop{5,kls}\cdop{3,k'l's'}|GS,4\rangle
\end{equation}
defined according to Eq. \ref{k_space_ph} and \ref{ph_ops}.

With this we can calculate the matrix element  
\begin{equation}
\langle kls,k'l's'|_{\text{p-h}}\,j^a(q=0)|GS\rangle=-\frac{\sqrt{2}}{3\hbar}(2s')^{2|l'|-1}
\sum_{\delta}\delta^a\sin(\vec{k}\cdot\vec{\delta})t^{0\,\delta}_{l,-l'}\,
\delta_{k,k'}\delta_{s,-s'}
\end{equation}

Introducing an explicit band structure, i.e. defining $t^{rr'}_{ll'}$, we can in principle calculate the 
optical conductivity due to the particle hole excitations. A lower bound
to the support of the sum in Eq. \ref{optcond} will tell us the optical gap, below which $\omega$, 
$\sigma(\omega)$ will decay. We simply maximize 
the kinetic energy of the particle and hole under the zero momentum constraint in the usual manner to obtain this. 
We will ignore the 
possible complications due to the dispersion and angular momentum part 
$\sum_{\delta}\delta^a\sin(\vec{k}\cdot\vec{\delta})t^{0\,\delta}_{l,-l'}$ which may kill the matrix 
elements at some high symmetry points. Moving slightly away from such symmetry points will give a finite contribution
to the response.

The optical gap is thus given by
\begin{equation}
\Delta_{\text{optical}}\geq U+4J_L+\frac{3}{2}J_S+(\frac{1}{3}\varepsilon^t_i(\vec{k})-
\frac{2}{3}\varepsilon^t_j(\vec{k}))_{\text{min}(\vec{k},i,j)}\,,\label{optgap}
\end{equation}
where $\varepsilon^t_j(\vec{k})$ is the kinetic energy of the $j$'th band at momentum $\vec{k}$ and 
$\text{min}(\vec{k},i,j)$ means minimizing with respect to the momentum and the band indices.

To get the Mott gap we just need to 
find the lowest energy particle-hole state, which will in general be smaller than the optical gap. 
This corresponds to putting the particle at the
bottom of the single particle band and the hole at the top 
\begin{equation}
\Delta_{\text{Mott}}=
U+4J_L+\frac{3}{2}J_S+(\frac{1}{3}\varepsilon^t_i(\vec{k})-
\frac{2}{3}\varepsilon^t_j(\vec{k'}))_{\text{min}(\vec{k},\vec{k'},i,j)}\,.
\end{equation}

\begin{figure}[h]
\includegraphics{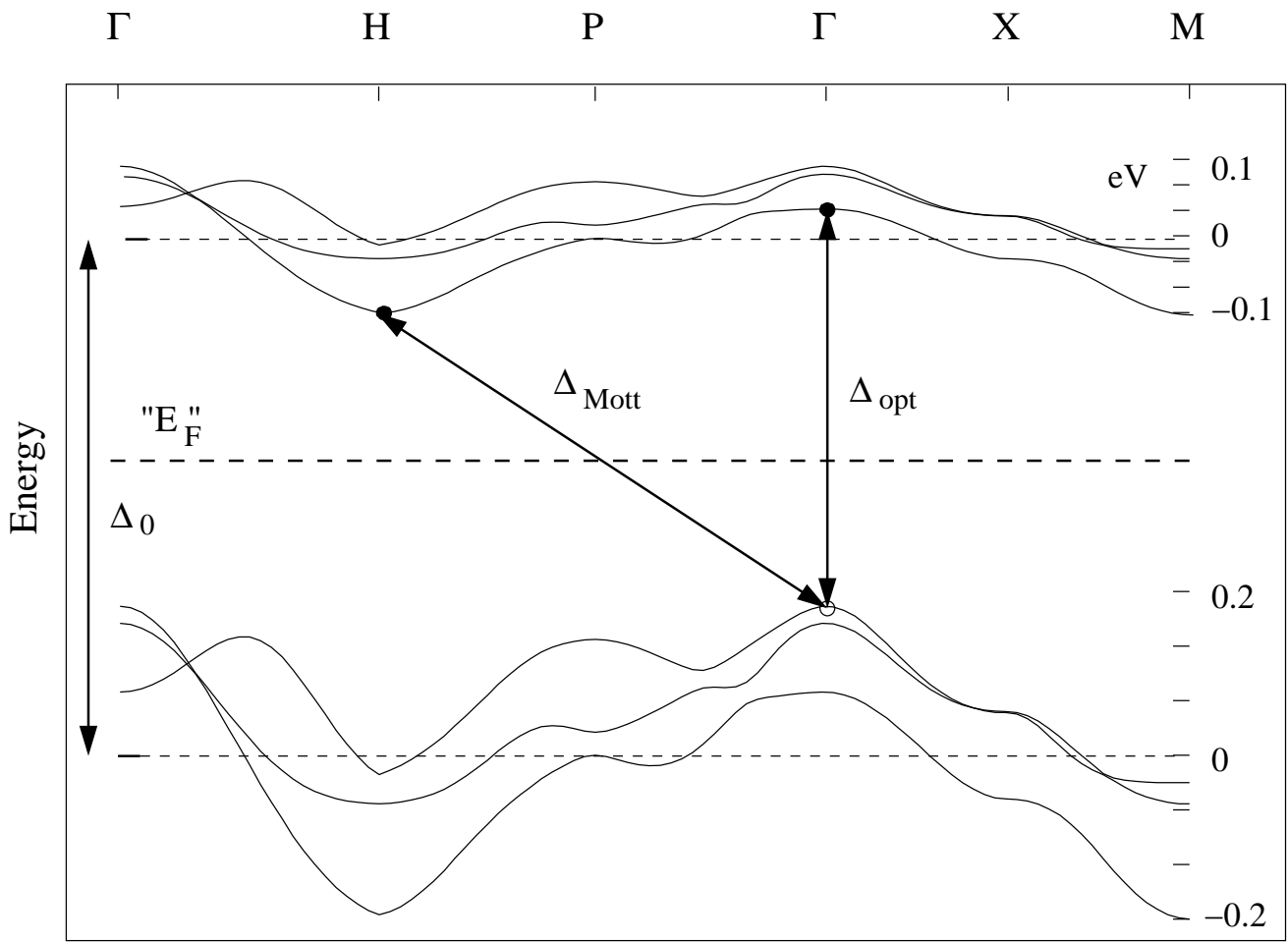}
\caption{\label{fig:chargegaps}Sketch of the band structure at doping $n=4$ 
(K$_4$\C60 \cite{GunnarssonErwin}) showing the indirect gap
structure with optical gap $\Delta_{\text{opt}}$ and Mott gap $\Delta_{\text{Mott}}$. The thin dashed lines
indicate the centers of the particle and hole bands and the energy scales on the right are defined with respect 
to these. $\Delta_0=U+4J_L+\frac{3}{2}J_S$ is the gap in $t=0$ limit.}
\end{figure}

In Figure \ref{fig:chargegaps} we present a caricature of the band structure around $n=4$ based on 
a an explicit non-interacting band structure calculated for K$_4$\C60 by Gunnarsson et al. 
\cite{GunnarssonErwin} which is rescaled by a factor $1/3$ for the particle band and $2/3$ for the hole band.
For this band structure, and any other where the max and min are not at the same k-vector, we find an indirect gap where
the optical gap is 
larger then the Mott gap and that the lowest energy particle-hole excitations have non-zero momentum.\cite{Halperin} 

A word of caution may be appropriate in considering this figure, namely that 
the single particle states in the bands are only well defined within our theory for a small density of such states. 
We depict the lower ``band'' as filled with
single particle states, but the real entities are only the holes in this band. 
This is a strongly interacting system
and the analogy with a weakly interacting semiconductor has limitations.  
For instance it is quite obviously nonsensical to fill up the particle band with a density of more than 
two particles because 
that would correspond to a total electron density of more than six. We will return to issue of doping away from 
the Mott insulator in section \ref{2+x}.

For comparison with other models of the Mott transition on degenerate Hubbard models it is useful to write down
a more general expression for the Mott gap.  If we make the reasonable assumption that the top and bottom of the 
band structure are roughly the same magnitude $W/2$,
where $W$ is the bandwidth of the tight binding Hamiltonian $h$, we can write 
\begin{equation}
\Delta_{\text{Mott}}\approx U_{eff}(4)-\half W\label{Mottgap}\,,
\end{equation}
where $U_{eff}(4)=U+4J_L+\frac{3}{2}J_S$.
This expression is in sharp contrast to calculations on $N$-band Hubbard models without the multiplet splitting 
terms, $\vec{S}^2$ and $\vec{L}^2$, where forms such as 
$\Delta_{\text{Mott}}\approx U-NW$ or $\Delta_{\text{Mott}}\approx U-\sqrt{N}W$ has been 
suggested.\cite{Lu_Gunnarsson_Kock} The intuitive motivation for the $N$ dependence is an increase in the kinetic
energy of the particle hole state due to the additional hopping channels. In this model
we see a different behavior since the number of hopping channels are limited by the strong spin dependent on-site
interactions.

\subsection{Why is A$_3$\C60 metallic?\label{sec:A3_metallic}}
One of the most striking facts about the fullerides is that the A$_3$\C60 materials are generally metallic given that 
the even integer filling materials are likely large $U$ insulators.   
The problem of odd-integer filling is significantly more complicated than that of even integer filling as presented
above. The reason for this is that even in the strong coupling limit, Eq. \ref{strong_coupling}, the ground state 
consists of states $(n=3,L=1,S=\half)$ with spin and orbital degeneracies.   
This problem resembles the half-filled single band Hubbard model with a highly degenerate
ground state which will be split to order $t^2/U$. The ground state may then have magnetic and/or
orbital order.  

However, if we
assume that the putative insulating ground state is not ordered so that the 
hopping of particles, $(n=4,L=0,S=0)$, and holes, $(n=2,L=0,S=0)$, is not  
frustrated by the spin interactions we can derive a Mott gap in analogous fashion to that for $n=4$ above
which reads
\begin{eqnarray}
\Delta_{\text{Mott}}(n=3)&\agt & U-4J_L-\frac{3}{2}J_S+(\frac{2}{3}\varepsilon^t_i(\vec{k})-
\frac{2}{3}\varepsilon^t_j(\vec{k'}))_{\text{min}(\vec{k}\vec{k'},i,j)}\nonumber\\
&\approx &U_{eff}(3)-\frac{2}{3}W\,,\label{Mottgapn=3}
\end{eqnarray}
where $W$ again is the non-interacting bandwidth of $h$. If the ground state has significant magnetic or orbital 
correlations we expect the gap to be bigger because of a lower ground state energy and frustration of the motion
of the particle and hole. 

Compared to the expression \ref{Mottgap} for the Mott gap at $n=4$ we note  
an increase from $W/2$ to $2W/3$ in the kinetic energy of the particle and holes due to the larger phase space 
allowed for hopping. In A$_3$\C60 where the bandwidth is around $.5eV$ it appears that this difference will not
be large enough to close the $.5eV$ Mott gap seen in A$_4$\C60.  

However, in addition there is in this model also a more distinct difference between even and odd integer filling,
namely the sign change of the $J_S$ and $J_L$ terms between the effective Hubbard
repulsion $U_{eff}(4)=U+4J_L+\frac{3}{2}J_S$ and 
$U_{eff}(3)=U-4J_L-\frac{3}{2}J_S$. This difference
comes from the fact that the lowest energy particle and hole excitations are $(L=0,S=0)$ at odd integer filling 
while they are $(L=1,S=\half)$ at even integer. If the multiplet splitting interactions are large enough they could  
certainly destabilize the Mott insulating ground state at odd integer filling. For instance, if $J_S$ and $J_L$ are 
very large 
such that $U_{eff}(3)<0$ and $|U_{eff}(3)|\mgt t$ we would have a spinless Bose liquid consisting of an 
equal number of two and four particle 
singlets which could only propagate to second order in $t$.\cite{Chakravarty_Kivelson} 
At intermediate coupling $U_{eff}(3)\sim t$  
we would expect some correlated metallic state with most of the spectral weight in low spin configurations of the 
two, three and four particle states, allowing for 
hopping to first order in $t$. Evidence for the formation of singlet configurations on short time scales
in the metallic fullerides Na$_2$Cs\C60, RB$_3$\C60 and the quenched cubic Cs\C60 ($n=1$)
have been presented from NMR spin-lattice relaxation measurements.\cite{BrouetII,BrouetIII}
  
In addition, we know that A$_3$\C60 is quite close to a metal insulator transition. 
It has been found that 
intercalating ammonia into the crystal can cause a transition into an insulating magnetically ordered 
phase.\cite{Takenedu} 
The main effect here is presumably the expansion of the lattice and corresponding decrease in the 
bandwidth, although the crystal symmetry is also reduced which may be important in facilitating a magnetically 
ordered ground state. An important consequence of these findings if interpreted through our model is  
that for A$_3$\C60 $U_{eff}(3)>0$ because the ground state for small $t$ is magnetic and 
that $|U_{eff}(3)|\sim t$ because changes in the magnitude of $t$ can induce a metal insulator transition.

\subsection{Experiment}
A basic observation from experiments is that there appears to be two distinct energy scales in these 
materials. Probes that are sensitive to spin, in particular NMR, are consistent with a spin gap of around 
50-100meV while optical conductivity sees a larger charge gap of around 500meV. Here we will look at the consistency
of the model with these observations and make a fit to estimate our microscopic parameters $U$, $J_L$ and $J_S$.

\subsubsection{Optical Gap}
The charge gap is seen in optical conductivity as a depletion of the low energy weight in K$_4$- and Rb$_4$\C60
below roughly 500meV.\cite{opt}
By inspection of Figure \ref{fig:chargegaps} together with the expression \ref{optgap} for the optical gap we
get the following estimate 
\begin{equation}
\Delta_{\text{opt}}(K_4C_{60})\approx U_{eff}(4)-200meV\approx 500meV\,,\label{opt_estimate}
\end{equation}
where again $U_{eff}(4)=U+4J_L+\frac{3}{2}J_S$. Solving Eq. \ref{opt_estimate} gives $U_{eff}(4)\approx 700meV$.

We can compare this to A$_3$\C60
where we expect the charge gap to close. Using expression \ref{Mottgapn=3} for the charge gap at $n=3$ together with
a bare bandwidth of $600meV$ gives
\begin{equation}
\Delta_{\text{Mott}}(A_3C_{60})\approx U_{eff}(3)-400meV\leq 0\,,
\end{equation}
with $U_{eff}(3)=U-4J_L-\frac{3}{2}J_S$. Together with the estimate $U_{eff}(3)>0$ as discussed in Sec. 
\ref{sec:A3_metallic} we thus find the rough estimate of 
$0<U_{eff}(3)\alt 400meV$. 

Combining the values for $U_{eff}(4)$ and $U_{eff}(3)$ we can now estimate the microscopic 
parameters of the model. We find $U\alt 550meV$ and 
$150meV\alt\frac{3}{2}J_S+4J_L<350meV$.

\subsubsection{NMR, $1/T_1$\label{NMR}}

Various probes\cite{triplet_evidence,Kerkoud} have detected a thermally activated magnetic susceptibility in 
K$_4$\C60 and Rb$_4$\C60 and more recently also in Na$_2$\C60\cite{Brouet}. This has 
been interpreted as evidence for a singlet-triplet gap of Jahn-Teller distorted molecules where a 
a molecule is thermally excited from the JT ground state singlet to the triplet which then
acts as a local moment. In the model presented here it is natural to assign such experimental signatures of 
gapped spin excitations to the spin modes or magnons which are a necessary part of the spectrum of the non-magnetic
Mott insulator. 

We will be focusing on measurements of $1/T_1$, the spin lattice relaxation rate, deriving the temperature dependence
of the relaxation by the magnons in the limit $T\mlt\Delta_s$.

The probability of a transition between nuclear spin states with $z$-component $m'=m\pm 1$ due to a two magnon 
process which scatters a magnon with $i=(L^z,S^z)$ and momentum $k$ to $i'=({L'}^z,{S'}^z)$ and $k'$ is given by
Fermi's golden rule as
\begin{equation}
W_{mm'}=\sum_{ik,i'k'}\frac{2\pi}{\hbar}|\langle m,n_{ik},{n'}_{i'k'}|V|m',n_{ik}-1,{n'}_{i'k'}+1\rangle|^2
\delta(E_{ik}-E_{i'k'})\,, 
\end{equation}
Here $n_{ik}$ is the magnon number operator, the interaction $V=A\vec{I}\cdot\vec{S}_0$ with hyperfine 
coupling $A$, nuclear spin $\vec{I}$ and electron spin at the nuclear
site $\vec{S}_0$. We have dropped the tiny, typically $\sim 10^{-6}eV$, Zeeman splitting of the nuclear spin.
 
The electron spin operator will act as a $b^{\dagger}_{r,{L'}^zS^z+(m-m')}b_{r,L^zS^z}$ with some small prefactor 
given by the overlap of these states with the single atom, where $b^{\dagger}_{r,L^zS^z}$ creates a 
triplet excitation. 
Ignoring the details of the matrix element between the different triplet states one finds that the 
general magnon matrix element is given by 
$\langle n_{i'k'}+1 \rangle\langle n_{ik} \rangle$, where $\langle n_{ik} \rangle=(e^{\beta E_{ik}}-1)^{-1}$ is just the
Bose occupation of the magnons. 
Finally the relaxation rate is given by \cite{Slichter} $\half\frac{\sum_{mm'}W_{mm'}(E_m-E_{m'})^2}{\sum_mE_m^2}$ 
where the magnon
part of $W_{mm'}$ clearly is independent of $m$ and $m'$. We thus arrive at the final expression 
\begin{equation} 
1/T_1\sim\int_{\Delta_s}^{\Delta_s+W_{\text{mag}}}
d\epsilon \frac{N^2(\epsilon)}{\sinh^2(\beta\epsilon/2)}
\end{equation}
where we have converted the sums to integrals by introducing the density of magnon states $N(\epsilon)$.
Here $\Delta_s$ is the spin gap, i.e. the lower edge of the magnon band, and $W_{\text{mag}}$ is the magnon band width. 

For $T<<\Delta_s$ we can replace the $1/\sinh^2(\beta\epsilon/2)$ by $4e^{-\beta\epsilon}$ and the integral is dominated
by $\epsilon\sim \Delta_s$. Assuming a quadratic dispersion at the band edge we get 
$N(\epsilon)\sim\sqrt{\epsilon-\Delta_s}$ for $\epsilon\agt\Delta_s$. By change of integration variable we arrive
at the temperature dependence 
\begin{equation}
1/T_1\sim T^2e^{-\Delta_s/T}\,,\quad T<<\Delta_s
\end{equation}

\begin{figure}[h]
\includegraphics{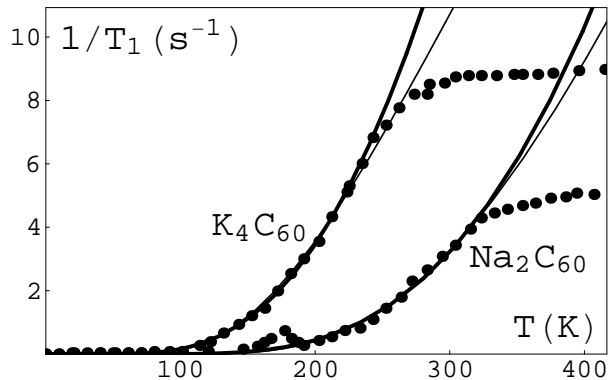}
\caption{\label{NMRfig}Fit to ${}^{13}$C NMR data by Brouet et al.\cite{Brouet}. The thick lines are fits to 
$1/T_1\sim T^2e^{-\Delta_s/T}$ with $\Delta_s=240K$ for K$_4$\C60 and $\Delta_s=710K$ for Na$_2$\C60. The thin
lines are fits to $1/T_1\sim e^{-\Delta_s/T}$ with corresponding $\Delta_s=660K$ and $1260K$. 
(The bump around 180K for Na$_2$\C60 is a presumably a molecular motion peak.)}
\end{figure}

Figure \ref{NMRfig} shows a fit of this model to ${}^{13}$C $1/T_1$ data on K$_4$\C60 and Na$_2$\C60.\cite{Brouet} 

We get an excellent fit to the below room temperature activated behavior with a value $\Delta_s=240K\approx 25meV$ 
for K$_4$\C60.
For comparison we also show fits to a model of localized triplet states corresponding to 
$W_{\text{mag}}=0$ and $1/T_1\sim e^{-\Delta_s/T}$. This is the fit used in the the experimental work 
\cite{triplet_evidence,Kerkoud,Brouet} which is based on a model of a static uniform Jahn-Teller singlet-triplet gap.   
In this intermediate temperature
regime it is difficult to tell which fit is best, in particular considering the fact that both models clearly fail
at higher temperatures where the relaxation rapidly saturates, and we conclude that the NMR data cannot resolve the two
scenarios.

Nevertheless, using the value $\Delta_s=25meV$ for the spin gap we may estimate the microscopic parameters. 
Given a bandwidth $W_{\text{mag}}$ which we
assume to be symmetric around the center we get 
\begin{equation}
\Delta_s=\dbarbar-W_{\text{mag}}/2\,,\label{spingap}
\end{equation}
where 
$\dbarbar=2J_L+2J_S$ is the $t=0$ spin gap. (There is an additional corrections to the spin gap of order 
$t^2/U$ which is a shift of the ground state energy which should be included in a more rigorous treatment.) 
A very rough estimate of the bandwidth may be given by the
degeneracy of the spin modes $W_{\text{mag}}=9t^2/U$. With $U\approx 500meV$ as derived from charge gap and taking
$t\approx 100meV$ form band structure calculations gives $W_{\text{mag}}\approx 200meV$. Collecting into Equation
\ref{spingap} for the spin gap gives $\dbarbar=2J_L+2J_S\approx 125 meV$ which seems in reasonable agreement with the 
estimate $150meV\alt\frac{3}{2}J_S+4J_L<350meV$ from the charge gap. 
This order of magnitude agreement for the coupling constants $J_S$ and $J_L$ is certainly encouraging in that it 
comes from experiments on two apparently separate 
physical quantities.

A fit to $1/T_1$ for Na$_2$\C60 gives a larger gap of around 700K. Within our model the $t=0$ spin gap $\dbarbar$ 
is the same for $n=2$ and $n=4$ so the differing spin gaps are somewhat unexpected. However, since the crystal 
structure is different, the tight binding 
Hamiltonians of these materials may be very different and consequently the bandwidth of the spin modes. In fact,
Na$_2$\C60 is fcc while K$_4$ and Rb$_4$ are body centered tetragonal. The natural interpretation for the variations
within this model is thus variations of the magnon bandwidth. Along the same lines we note 
the behavior of $1/T_1$ in Rb$_4$\C60 under pressure where it is found that the activated behavior is replaced
or coexists with a non-activated component related to gapless excitations.\cite{Kerkoud} 
It has been suggested that this is related to a closing of the Mott gap due to the expected pressure induced 
increase of the bare bandwidth. Within our model we find a possible alternative interpretation in 
terms of a closing of the spin gap.

Above room temperature the activated behavior stops and the the relaxation rate saturates. Within our simple 
non-interacting model for the spin modes we cannot expect to be able to address the high-temperature behavior when 
a significant number of modes are excited. A more sophisticated treatment requires us to properly account 
for the interactions between the spin modes as well as the exclusion statistics that are ignored in the single 
particle picture. The saturation could also be related to molecular degrees of freedom at higher temperatures which 
are completely neglected in our model.\cite{Brouet}
However, we note that this rapid saturation is very reminiscent of the behavior of $1/T_1$ in 
spin ladder materials with gapped magnons\cite{Ladder_magnons}, 
where it is believed to have a purely electronic origin\cite{Lee}.

\section{The doped Mott insulator\label{2+x}}
We will now look at the problem of an incommensurate particle density away from the Mott insulators at even integer
filling $n=2$ and $n=4$. This is obviously a much more difficult task because in the strong coupling limit the ground 
state will be highly degenerate. As a concrete example, at a doping $n=2+x$ ($x<1$) the ground state at $t=0$ is the 
set of 
states with $x$ 3-particle states $(3,1,\half)$ and $(1-x)$ 2-particle states $(2,0,0)$ at arbitrary positions in 
space. Introducing $t$ by means of the the effective Hamiltonian (\ref{Hefffinal}) we find to first order
in $t$ nearest neighbor interchange of the 2- and 3-particle states and to second order in $t$ spin and orbital 
exchange terms between nearest neighbor 3-particle states. This can be described by a generalized $t-J$-model  
including a no double occupancy constraint because only 2 and 3 particle states 
are allowed. 

At first glance this may appear to be an even more difficult problem than that of a doped 
antiferromagnet because of the additional orbital degrees of freedom.
However, in the low density limit, $x\mlt 1$, it is in fact considerably simpler than the doped 
antiferromagnet because only the doped particles (or holes) have internal spin and 
orbital degrees of freedom. In the doped antiferromagnet the scenario is just the opposite with a large number $1-x$
of spinful particles and a small number $x$ of spinless holes. This of course gives rise to the very complex behavior
in such systems where the spin interactions $J$ can compete with the hopping $t$ even in the limit $J\sim t^2/U\mlt t$
because the important energetics is given roughly by $xt$ and $(1-x)J$. 
Here for the doped non-magnetic Mott insulator a similar consideration would lead us to compare $xJ$ with $xt$   
because it is the doped particles or holes that carry both the spin and momentum. Effectively we are thus looking 
at the low density (heavily doped) limit of a t-J model. 

We will completely neglect the nearest neighbor exchange interactions as well as the no double occupancy 
constraint and only consider the single particle physics. Certainly, for the problem of a single particle or hole 
doped into the non-magnetic Mott insulator this is completely rigorous and again in sharp contrast to the problem 
of a single hole in an antiferromagnet where interactions obviously cannot be neglected. Even this single particle 
physics has some interesting implications for the doping dependence in the metallic fullerides.  

\subsection{Small Fermi surface}

The problem of a single particle or hole was addressed already in Section \ref{single_particle_section} in connection
with particle-hole excitation. There we showed that the single particle or hole spectra are equivalent to 
the non-interacting $H_I=0$ spectrum up to a rescaling by a factor $1/3$ or $2/3$. 
At least in the very low density limit, $n=2+x$ or $n=4+x$ with $|x|\mlt 1$, we expect these single particle states
to give a qualitatively accurate picture by filling $x$ such states. 

In particular this implies a ``small Fermi surface'' where the number of delocalized charge carriers 
is proportional to the number of doped holes or particles $x$ and not the total filling $n$. 
The remaining degrees of freedom are frozen below the Mott gap. One important consequence is that the density of states 
at the Fermi surface for small doping $x$ will be given by the density of states at the band edges of the 
of non-interacting problem by the simple relation 

\begin{eqnarray}
DOS_{\text{strong coupling}}(n=2-x)& \approx &3 DOS_{\text{non-interacting}}(n=6-x) \nonumber\\
DOS_{\text{strong coupling}}(n=2+x)& \approx &3/2 DOS_{\text{non-interacting}}(n=x) \nonumber\\
DOS_{\text{strong coupling}}(n=4-x)& \approx &3/2 DOS_{\text{non-interacting}}(n=6-x) \nonumber\\
DOS_{\text{strong coupling}}(n=4+x)& \approx &3 DOS_{\text{non-interacting}}(n=x),\quad x\mlt 1.\label{small_x_DOS}
\end{eqnarray}

%

\begin{figure}[h]
\includegraphics{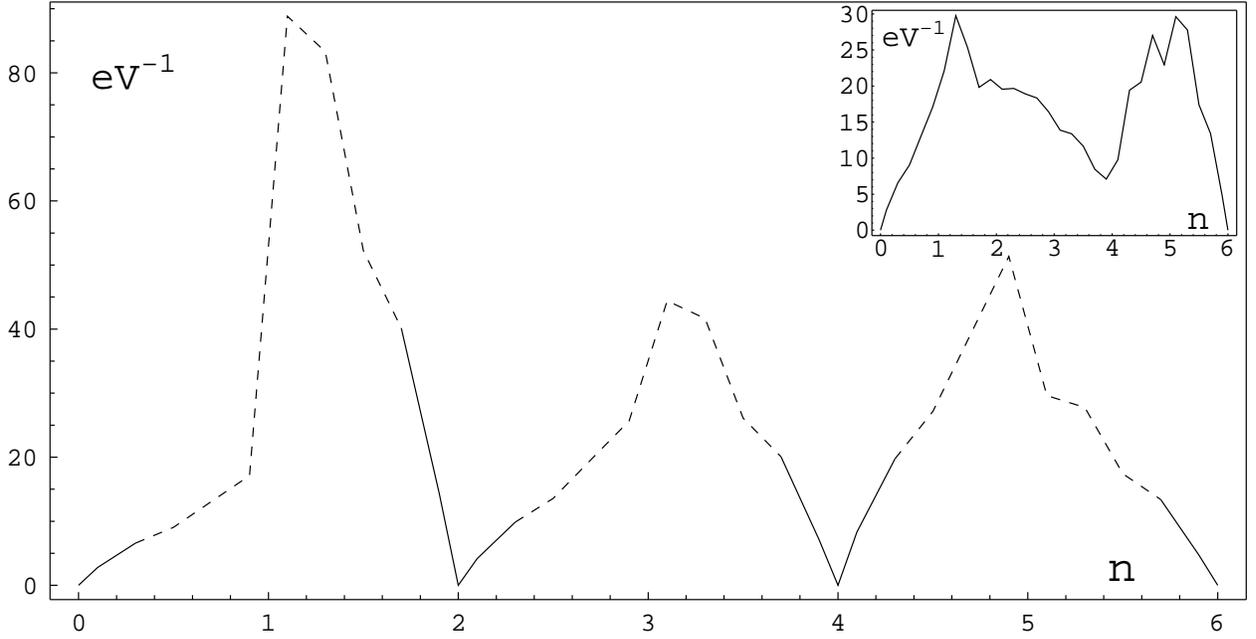}
\caption{\label{dosfig}Density of states (both spins) as function of filling $n$ given the 
non-interacting DOS inset. The dashed lines are extrapolations of Eqn \ref{small_x_DOS} toward odd integer filling.
The non-interacting DOS is calculated from the band structure of unidirectional A$_3$\C60.\cite{Laouini}}
\end{figure}

A detailed picture of what happens at larger doping $x\rightarrow 1$ as we approach odd integer filling is 
beyond our methods. 
However, a naive extrapolation of the results valid for small $x$ all the way to $x=1$ gives a 
density of states as shown in Figure \ref{dosfig}, where the density of states is generally peaked at 
odd integer filling as a consequence of the rapid decay toward the effective band edges at even integer filling.
For reasons discussed in Section \ref{sec:A3_metallic} we have to be in an intermediate coupling regime where 
the system is metallic at odd integer filling for this extrapolation to have any credibility.

There is an interesting experiment that corroborates the small Fermi surface picture in \C60 which is
the variation of density of states in Na$_2$Cs$_x$\C60 ($0<x<1$) corresponding to a doping range 
$2<n<3$.\cite{Yildrim}.
It was estimated from the Pauli susceptibility that for samples with $n=2.25,2.5,2.75,3$ the density of states 
varies as $5,7,11,15$ $eV^{-1}$ (both spins).  
Correspondingly, T$_c$ drops rapidly from 12K at $n=3$ to 7K at $n=2.75$ and to $<0.5K$ at $n=2.5$.
Certainly, this behavior seems consistent with the scenario sketched in figure \ref{dosfig} where the density of states
drops rapidly as the effective Hubbard band edges are approached at even integer filling.

\begin{figure}[h]
\includegraphics{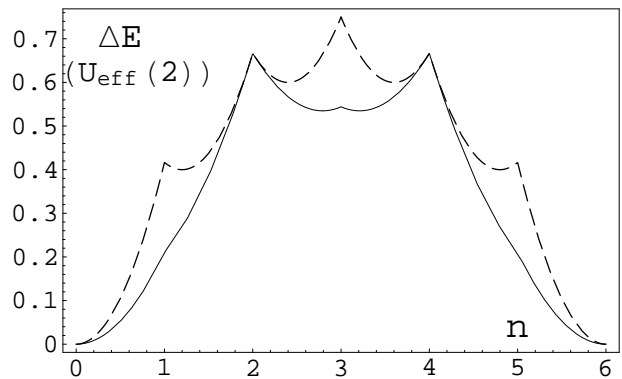}
\caption{\label{small_vs_large}Potential energy difference, in units of
$U_{eff}(2)=U+4J_L+3/2J_S$, between an uncorrelated and correlated ground state as a function of filling $n$. 
The solid line is for $J_S=2J_L=.1U$ and the dashed line for$J_L=J_S=0$.}
\end{figure}

The failure of naive band theory in the 
presence of strong local repulsion is a consequence of the 
large inherent charge fluctuations of such an uncorrelated state of delocalized electrons. 
It may be illuminating to recall some of Hubbard's original work on the topic of narrow band
systems.\cite{Hubbard}
For an $m$-fold degenerate band at filling $n$ the probability $P_N(m,n)$ of having $N$ 
electrons on a particular atom (molecule) is given by 
\begin{equation}
P_N(m,n)=\binom{m}{N}(\frac{n}{m})^N(1-\frac{n}{m})^{m-N}\,,\label{pn}
\end{equation}
where $\binom{m}{N}$ is the multiplicity of atomic states with $N$ particles. 
The rms fluctuation
is given by $(\Delta_N)_{\text{RMS}}=\sqrt{n(1-n/m)}$ which has a maximum $(\Delta_N)_{\text{RMS}}=\sqrt{m}/2$ 
at half filling $n=m/2$. Clearly, at finite doping there are significant charge fluctuations of order one which 
cost an energy of order $U$ per site in an uncorrelated state and which grows with the degeneracy. 
We can get an estimate of the energy cost of the 
charge fluctuations for our model by comparing the potential energy $E=\langle H_I\rangle$ in an uncorrelated state 
which is the 
ground state of the kinetic energy $h$ with that given by the small Fermi surface state which is the ground state of the 
potential energy $H_I$. The latter
is at filling $n$ given by $p_N$ N-particle states and $p_{N+1}$ $N+1$-particle states in the lowest energy multiplet 
where $N\leq n<N+1$ and $p_N N + p_{N+1}(N+1)=n$. The potential energy of this state is given by  
\begin{equation}
\langle H_I\rangle_{\text{corr}}=p_N E_0(N)+p_{N+1}E_0(N+1)\,,
\end{equation}
where $E_0(N)$ is the energy of the ground state multiplet 
with $N$ particles. 
The expression (\ref{pn}) turns into
\begin{equation}
P_{N,L,S}(n)=(2L+1)(2S+1)(\frac{n}{6})^N(1-\frac{n}{6})^{6-N}
\end{equation}
for this model where the $N$-particle multiplets are split
according to $L$ and $S$ and the corresponding potential energy for the uncorrelated state is 
\begin{equation}
\langle H_I\rangle_{\text{uncorr}}=\sum_{N,L,S}P_{N,L,S}(n)E(N,L,S)\,.
\end{equation}
Figure \ref{small_vs_large} shows 
$\Delta E=\langle H_I\rangle_{\text{uncorr}}-\langle H_I\rangle_{\text{corr}}$ as a function of doping and in units of 
$U_{eff}(2)=U_{eff}(4)=U+4J_L+3/2J_S$ for two cases $J_L=J_S=0$ and $J_L=.1U$, $J_S=.2U$.   

We see that in a wide doping range around half filling there is a significant energy cost due to charge fluctuations 
in an uncorrelated state. Given that this leads to the Mott insulating behavior at even integer filling we
would also expect significant correlation effects in the metallic regions $2<n<4$, consistent with the small Fermi surface
scenario.

Hubbard's treatment of narrow band systems used a real space Greens function method which is exact in the 
zero bandwidth $t=0$ case but 
which depend critically on neglecting correlations between electrons
on different atoms in the finite bandwidth case. He found that in the narrow bandwidth limit the original 
non-interacting band splits into a
large number of bands which correspond to transitions between states with particle number differing by one. 
In addition these Hubbard bands reflect the non-interacting band with density of states which are some functional of 
the non-interacting density of states. 
Our result for the doped Mott insulator is essentially a special case of Hubbard's results for which we can solve 
for the quasi particle spectrum exactly in the low density 
limit where we can neglect interactions between the doped particles or holes .

\subsection{High Temperatures}
In the previous section we discussed the ground state properties of the doped Mott insulator close to even integer 
filling. We found a band of single particle states which up to a rescaled hopping are equivalent to the states 
of the non-interacting problem. At low 
temperatures we thus expect a simple metallic behavior with band like charge transport. Here we will speculate on some 
of the interesting physics which could emerge from the model at higher temperatures.

The derivation 
of the single particle Hamiltonians, Eq. \ref{5sp} and  \ref{3sp}, for the particle and hole states depend 
crucially on the 
fact the Mott insulating ground state is of the simple non-degenerate form  $|GS,4/2\rangle=\prod_{r}|n=4/2,L=0,S=0>_r$ 
and that we can ignore the higher orbital and spin multiplets in the 2 and 4 particle molecular spectra. 
At elevated temperatures of order the spin gap $\Delta_s$ this assumption is no longer justified as the the spin and 
angular momentum modes discussed in Section \ref{sec:spin_modes} are thermally occupied and we should consider 
their effect on the quasiparticle states. (The activated behavior discussed in Sec. \ref{NMR} which has been  
linked to gapped triplet excitations is seen in NMR ($1/T_1$) also in the metallic Na$_2$Cs\C60 and 
Rb$_3$\C60.\cite{BrouetII})
These, in fact, interact very strongly with the quasiparticles in a quite 
non-trivial fashion.
In the presence of such an excited spin or orbital state the nearest neighbor hopping integrals described by Equation 
\ref{hopping_integrals} turn into some more complicated expression given by for instance 
\begin{equation}
\langle 4,1,1,{L'}^z,{S'}^z|_r\langle 3,l',s'|_{r'}
(\sum_{jj',\sigma}t^{rr'}_{j'j}c^{\dagger}_{r,j,\sigma}c_{r',j',\sigma})
|3,l,s\rangle_r|4,1,1,{L}^z,{S}^z\rangle_{r'}\,,
\end{equation} 
in the case of a hole hopping to a site occupied by a spin triplet. 
Depending on the configurations of the spin triplet states the hopping of the hole ($|3,l,s\rangle$) may be completely 
suppressed or it may require a spin flip. Certainly the nearest-neighbor hopping integrals would be completely altered
from the simple form of Equation \ref{hopping_integrals}.  

It is not obvious how to model this problem but the most naive scenario might be to
ignore the hopping of the spin and orbital states and replace them by thermally excited impurities 
causing disorder in the hopping of the particles or holes.    
Such a model would be similar to that suggested by Varma to explain the paramagnetic insulator to ferromagnetic metal 
transition in lanthanum manganites (giant magnetoresistive compounds).\cite{Varma} 

One may speculate that this temperature activated off-diagonal disorder could destroy the band like motion of the 
charge carriers and possibly be related to anomalous properties of A$_3$\C60
at elevated temperatures such as the evidence for localization from NMR\cite{BrouetII}, the disappearance of the Fermi 
edge\cite{Knupfer} and the non-saturation of 
resistivity and the correspondingly extremely short mean free paths\cite{Hebard}.
 
\section{Conclusions\label{Summary}}
We have studied an orbitally degenerate 3-band Hubbard model with additional multiplet 
splitting on-site interactions $J_S\vec{S}^2$ and $J_L\vec{L}^2$ which 
favor low spin and low orbital angular momentum. We use the effective Hamiltonian method in the strong coupling limit
$U\mgt J_S\mgt J_L\mgt t$ perturbatively to second order in $t$.  
At even integer filling, $n=2$ or $n=4$, this model is a insulator with a non-degenerate ground state where the 
electrons at each site occupy the $L=0$ and $S=0$ configurations and with distinct spin and charge gaps.  
The trivial ground state allows for a simple single particle description of spin and charge excitations.
The lowest energy spinful excitations is a band of magnons with a bandwidth $W_{\text{mag}}\sim t^2/U$ and a gap
$\Delta_s=2J_L+2J_S-\sO(t^2/U)$.
A single particle or hole doped into the Mott insulator are described by the non-interacting tight-binding Hamiltonian
but with an overall rescaling by a factor $1/3$ or $2/3$ of the hopping integrals and corresponding bandwidth.
The latter allows for a detailed description of the particle-hole excitations and the corresponding charge gap 
is given by $\Delta_{\text{Mott}}\approx U+4J_L+3/2J_S-W/2$ in terms of the 
band width $W$ of the non-interacting Hamiltonian. 

Close to the Mott insulator, at filling $2+x$ or $4+x$ with $|x|\mlt 1$, we find a metallic state with a ``small
Fermi surface'' where the density of charge carriers is given by $|x|$ and a density of states
which is simply a renormalization by a factor $3$ or $3/2$ of the density of states at the band edges 
of the non-interacting 
band structure. Consequently, in three dimensions the density of states will in general increase 
rapidly with the the doping $x$.

In this model there is also a distinct difference between even and odd integer filling which follows from the simple 
fact that an odd number of electrons cannot form a spin singlet.      
From this follows that the effective on-site repulsion is given by $U_{\text{eff}}(n=2/4)=U+4J_L+3/2J_S$ at 
even integer filling
and by $U_{\text{eff}}(n=1/3/5)=U-4J_L-3/2J_S$ at odd. 
Consequently, depending on the magnitude of $J_L$ and $J_S$, the Mott gap may be significantly reduced or vanish
at odd integer filling.   

The properties of this model are strikingly similar to the phenomenology of the fullerides A$_n$\C60 with 
$2\leq n\leq 4$. The non-magnetic Mott insulator at even integer filling with a small spin gap and a larger 
charge gap, the even/odd effect at integer doping where A$_3$\C60 is generally metallic, as well as
the rapid suppression of the DOS and the corresponding superconducting transition temperatures as the 
filling approaches 
even integer. We do a fit of the model to the charge gap from optical conductivity and the spin gap from 
NMR $1/T_1$ in K$_4$\C60 which appear consistent with values of $J_S$ and $J_L$ of around $50-100meV$.

There is a number of interesting open questions about the model and the possible implications to alkali doped 
\C60.
In particular we need a better understanding of the physics at odd integer filling on the metallic side of the
Mott transition.  
Can this state have a superconducting ground state even though, as evidence suggest, $U_{eff}(3)>0$ 
such that there is no bare attraction in the way envisioned by Chakravarty and coworkers as an electronic 
mechanism of superconductivity? In fact, also at even integer filling there have been intriguing suggestions of an 
intermediate superconducting state in the metal-insulator transition.\cite{CaponeII,CaponeIII}
Another issue is the properties of the model at elevated temperatures approaching the spin gap where we have found that 
the spin and orbital modes interact strongly with the charge carries and may 
significantly effect the simple band like charge transport.

\appendix*

\section{Second order canonical transformation.}
Here we derive the expressions for the second order terms in the effective Hamiltonian Eq. \ref{Hefffinal}.
Starting with Eq. \ref{Heff1}
\begin{equation}
\sH_{eff}=H_I+h^0+i[\sS_1,h^0]+\frac{i}{2}[\sS_1,h^1]+i[\sS_2,H_I]\nonumber\,,
\end{equation}
where $\sS_1$ is given by 
\begin{equation}
\sS_1=\sum_{\langle r,r'\rangle}\frac{ih^{1,rr'}_{ab}}{E(b)-E(a)}|a\rangle_{rr'}\langle b|_{rr'}
\end{equation}

Our purpose is to construct $\sS_2$ such that it cancels terms which connect different energy subsectors from the 
commutators $[\sS_1,h^{0/1}]$.  
We have 
\begin{eqnarray}
i[\sS_1,h^{0/1}]&=&-\sum_{\langle r,r'\rangle,\langle r'',r'''\rangle}\frac{h^{1,rr'}_{ab}}{E(b)-E(a)}
h^{0/1,r''r'''}_{cd}[|a\rangle_{rr'}\langle b|_{rr'},|c\rangle_{r''r'''}\langle d|_{r''r'''}]\nonumber\\
&=&-\sum_{\langle r,r'\rangle}\frac{h^{1,rr'}_{ab}}{E(b)-E(a)}
h^{0/1,rr'}_{cd}[|a\rangle_{rr'}\langle b|_{rr'},|c\rangle_{rr'}\langle d|_{rr'}]\nonumber\\
&&-\sum_{\langle r,r',r''\rangle}\frac{h^{1,rr'}_{ab}}{E(b)-E(a)}
h^{0/1,r'r''}_{cd}[|a\rangle_{rr'}\langle b|_{rr'},|c\rangle_{r'r''}\langle d|_{r'r''}]\,,
\end{eqnarray}
where we have used the fact the operators commute if there is no overlap between sites.
Using the complete set of three site states $1=\prod_a|a\rangle_{rr'r''}\langle a|_{rr'r''}$ for the three site 
interaction we arrive at 
\begin{eqnarray}
i[\sS_1,h^{0/1}]&=&-\sum_{\langle r,r'\rangle,c}\left(\frac{h^{1,rr'}_{ac}}{E(c)-E(a)}
h^{0/1,rr'}_{cb}+h^{0/1,rr'}_{ac}\frac{h^{1,rr'}_{cb}}{E(c)-E(b)}\right)|a\rangle_{rr'}\langle b|_{rr'}\nonumber\\
&-&\sum_{\langle r,r',r''\rangle,c}\left(\frac{h^{1,rr'}_{ac}}{E(c)-E(a)}
h^{0/1,r'r''}_{cb}+h^{0/1,r'r''}_{ac}\frac{h^{1,rr'}_{cb}}{E(c)-E(b)}\right)|a\rangle_{rr'r''}\langle b|_{rr'r''}
\end{eqnarray}

We now split this into a part $\delta_{E(a)E(b)}$ which is diagonal in energy and $(1-\delta_{E(a)E(b)})$
which connects different energy sectors. The latter part we can cancel by solving for $\sS_2$,
schematically
\begin{equation}
[\sS_2,H_I]=-([\sS_1,h^0]+\frac{1}{2}[\sS_1,h^1])_{\text{off-diagonal}}\,.
\end{equation}
Using an ansatz $\sS_2=\sum_{\langle r,r'\rangle}\sS_{2,ab}^{rr'}|a\rangle_{rr'}\langle b|_{rr'}+
\sum_{\langle r,r',r''\rangle}\sS_{2,ab}^{rr'r''}|a\rangle_{rr'r''}\langle b|_{rr'r''}$ and  
using the fact that $[|a\rangle_{rr'r''}\langle b|_{rr'r''},H_I]=(E(b)-E(a))|a\rangle_{rr'r''}\langle b|_{rr'r''}$,
we can solve for $\sS_2\sim\sO(t^2/\Delta E)$. Apart from the cancellation, $\sS_2$ will only contribute to 
higher orders.

Having done the cancellation we are left with terms that are diagonal in energy. Clearly $[\sS_1,h^0]$ does 
not contribute to this because $h^0$ is diagonal in energy and $\sS\sim h^1$ is strictly off-diagonal. The result
for the remaining second order terms is 
\begin{eqnarray}  
\frac{i}{2}[\sS_1,h^{1}]_{\text{diagonal}}&=&-\half\sum_{\langle r,r'\rangle,c}\frac{h^{1,rr'}_{ac}h^{1,rr'}_{cb}
+h^{1,rr'}_{ac}h^{1,rr'}_{cb}}{E(c)-E(a)}
\delta_{E(a)E(b)}|a\rangle_{rr'}\langle b|_{rr'}\nonumber\\
&-&\half\sum_{\langle r,r',r''\rangle,c}\frac{h^{1,rr'}_{ac}
h^{1,r'r''}_{cb}+h^{1,r'r''}_{ac}h^{1,rr'}_{cb}}{E(c)-E(a)}\delta_{E(a)E(b)}|a\rangle_{rr'r''}
\langle b|_{rr'r''}\,,
\end{eqnarray}
which simplifies to the final expression given in \ref{Hefffinal}.


\begin{thebibliography}{33}
\bibitem{Sachdev} S. Sachdev, Annals Phys. {\bf 303}, 226 (2003). 
\bibitem{Mott_Note} We use the term Mott insulator in the broad context of an insulator which according to band theory 
calculations should be a metal because there is at least one partially filled band. However, we will 
primarily consider even 
integer filling which could correspond to a band insulator because there is an even number of
electrons per unit cell. A more distinct definition of a Mott insulator valid in our case is an insulator 
in which the electronic charge gap and spin gap are different.
\bibitem{Gunnarsson_review} For a review see O. Gunnarsson, Rev. Mod. Phys {\bf 69}, 575 (1997).
\bibitem{opt} Y. Iwasa and T. Kaneyasy, Phys. Rev. B {\bf 51}, 3678 (1995), 
M. Knupfer and J. Fink, Phys. Rev. Lett. {\bf 79}, 2714 (1997).
\bibitem{triplet_evidence} I. Lukyanchuk, N. Kirova, F. Rachdi, C. Goze, P. Molinie, and M. Mehring, Phys. Rev. B {\bf 51}, 3978 (1995), 
G. Zimmer, M. Mehring, C. Goze, and F. Rachdi, Phys. Rev. B {\bf 52}, 13300 (1995).
\bibitem{Kerkoud} R. Kerkoud, P. Auban-Senzier, D. J\'erome, S. Brazovskii, I. Lukyanchuk, N. Kirova, F. Rachid, and C. Goze, J. Phys Chem Solids, {\bf 57}, 143 (1996).
\bibitem{Yildrim} T. Yildirim, L. Bardette, J.E. Fischer, C.L. Lin, J. Robert, P. Petit, and T.T.M. Palstra , Phys. Rev. Lett. {\bf 77}, 167 (1996).
\bibitem{Auerbach} N. Manini, E. Tosatti and A. Auerbach, Phys. Rev. B {\bf 49}, 13008 (1994).
\bibitem{Kuntscher} C.A. Kuntscher, G.M. Bendele and P.W. Stephens, Phys. Rev. B {\bf 55}, 3366 (1997).
\bibitem{Chakravarty} S. Chakravarty, M.P. Gelfand and S.A. Kivelson, Science {\bf 254}, 970 (1991).  
\bibitem{Baskaran} G. Baskaran and E. Tosatti, Curr. Sci. {\bf 61}, 33 (1991). 
\bibitem{antiHund} For a discussion of Hund's rules and violations of them, see 
W. Kutzelnigg and J.D. Morgan III, Z. Phys. D {\bf 36}, 197 (1996). 
\bibitem{White} S.R. White, S. Chakravarty, M.P. Gelfand and S.A. Kivelson, Phys Rev. B {\bf 45}, 5062 (1992).    
\bibitem{Fabrizio} M. Fabrizio and E. Tosatti, Phys. Rev. B {\bf 55}, 13465 (1997).
\bibitem{CaponeI} M. Capone, M. Fabrizio, P. Giannozzi, and E. Tosatti, Phys. Rev. B {\bf 62}, 7619 (2000). 
\bibitem{CaponeII} M. Capone, M. Fabrizio, and E. Tosatti, Phys. Rev. Lett. {\bf 86}, 5361 (2001). 
\bibitem{CaponeIII} M. Capone, M. Fabrizio, C. Castellani, and E. Tosatti, Science {\bf 296}, 2364 (2002). 
\bibitem{Luders} M. L\"{u}ders, A. Bordoni, N. Manini, A. DalCorso, M. Fabrizio, and E. Tosatti, Philos. Mag. B {\bf 82}, 1611 (2002). 
\bibitem{Lemmert} P.E. Lemmert, D.S. Rokhsar, S. Chakravarty, S. Kivelson, and M.I. Salkola, Phys. Rev. Lett. {\bf 74}, 996 (1995).
\bibitem{Laouini} N. Laouini, O.K. Andersen and O. Gunnarsson, Phys. Rev. B {\bf 51}, 17446 (1995).
\bibitem{Fazekas} P. Fazekas, {\em Lecture Notes on Electron Correlation and Magnetism} (World Scientific Singapore) 1999. 
\bibitem{Mahan} G.D. Mahan, {\em Many-Particle Physics} 2nd ed. (Plenum Press New York) 1990. 
\bibitem{GunnarssonErwin} O. Gunnarsson, S.C. Erwin, E. Koch and R.M. Martin, Phys. Rev. B {\bf 57}, 2159 (1998). 
\bibitem{Lu_Gunnarsson_Kock} J.P. Lu, Phys. Rev. B {\bf 49}, 5687 (1994), 
O. Gunnarsson, E. Koch and R.M. Martin, Phys. Rev. B {\bf 54}, 11026 (1996).
\bibitem{Halperin} For a review on semiconductors with indirect gaps, see B.I. Halperin and T.M. Rice, 
in Solid States Phy. V.21 (1968). (Seity, Tumball and Ehrenreich, eds.) 
\bibitem{Chakravarty_Kivelson}  S. Chakravarty and S.A. Kivelson, Phys. Rev. B {\bf 64}, 064511 (2001).
\bibitem{BrouetII} V. Brouet, H. Alloul, S. Garaj and L. Forr\'o, Phys. Rev. B {\bf 66}, 155124 (2002).
\bibitem{BrouetIII} V. Brouet, H. Alloul and L. Forr\'o, Phys. Rev. B {\bf 66}, 155123 (2002).
\bibitem{Takenedu} Y. Iwasa, H. Shimoda, T.T.M. Palstra, Y. Maniwa, O. Zhou, and T. Mitani, Phys. Rev. B {\bf 53}, 8836 (1996), 
T. Takenobu T. Muro, Y. Iwasa, and T. Mitani, Phys. Rev. Lett. {\bf 85}, 381 (2000).
\bibitem{Brouet} V. Brouet, H. Alloul, L. Thien-Nga, S. Garaj and L. Forr\'o, Phys. Rev. Lett. {\bf 86}, 4680 (2001),
V. Brouet, H. Alloul, S. Garaj and L. Forr\'o, Phys. Rev. B {\bf 66}, 155122 (2002). 
\bibitem{Slichter} C.P. Slichter, {\em Principles of Magnetic Resonance} (Springer New York) 1990.
\bibitem{Ladder_magnons} T. Imai, K.R. Thurber, K.M. Shen, A.W. Hunt, and F.C. Chou, Phys. Rev. Lett. {\bf 81}, 220 (1998), 
Y. Piskunov, D. J\'erome, P. Auban-Senzier, P. Wzietek, C. Bourbonnais, U. Ammerhal, G. Dhalenne, and A. Revcolevschi, Euro. Phys. J. B {\bf 24}, 443 (2001).
\bibitem{Lee} D.A. Ivanov and P.A. Lee, Phys. Rev. B {\bf 59}, 4803 (1999).
\bibitem{Hubbard} J. Hubbard, Proc. R. Soc. London, A {\bf 276}, 238 (1963), {\em ibid.} A {\bf 277}, 237 (1964).
\bibitem{Varma} C.M. Varma, Phys. Rev. B {\bf 54}, 7328 (1996). 
\bibitem{Knupfer} M. Knupfer, M. Merkel, M.S. Golden, J. Fink, O. Gunnarsson, and V.P. Antropov, Phys. Rev. B {\bf 47}, 13944 (1993)
\bibitem{Hebard} A.F. Hebard, T.T.M. Palstra, R.C. Haddon and R.M. Fleming, Phys. Rev. B {\bf 48}, 9945 (1993).
\end{thebibliography}

\end{document}